\documentstyle[11pt,newpasp,twoside,epsf]{article}
\def\edcomment#1{\iffalse\marginpar{\raggedright\sl#1\/}\else\relax\fi}
\reversemarginpar

\begin{document}
\title{High resolution spectral survey of symbiotic stars in the near-IR over
the GAIA wavelength range}
\author{Paola M. Marrese$^{1,2}$, Rosanna Sordo$^{2}$ and Ulisse Munari$^{1}$ }
\affil{1)~~Padova Astronomical Observatory INAF, Asiago station, via Osservatorio 8, I-36012 Asiago (VI), Italy}
\affil{2)~~Department of Astronomy, University of Padova, vicolo Osservatorio 8, I-35122 Padova, Italy}
\begin{abstract}
High resolution (R$\sim$20,000), high signal-to-noise (S/N $\geq$ 100) spectra were collected for $\sim$40 
symbiotic stars with the Asiago echelle spectrograph~ over the same 8480-8740 \AA~ wavelength range 
covered by the ESA Cornerstone mission GAIA, centered on the near-IR CaII triplet and the head of 
the Paschen series. A large number ($\sim$ 140) of cool MKK giant and supergiant templates 
were observed with the same instrumentation to serve as a reference and classification grid.\\
The spectra offer bright prospects in classifying and 
addressing the nature of the cool component of symbiotic stars (deriving T$_{{\sl eff}}$, 
log{\sl g}, [Fe/H], [{\sl $\alpha$}/Fe], V$_{{\sl rot}}$sin{\sl i} both via MDM-like methods and syntetic atmosphere 
modeling) and mapping the physical condition and kinematics of the gas regions responsible for the emission 
lines.
\end{abstract}
\section{Introduction}
The spectral region around the near-IR CaII triplet and the head of the Paschen series is among the most 
promising to investigate the cool component of symbiotic stars. First of all the region is free from 
telluric absorptions and thus well accessible from the ground (Munari 1999). Secondly, given the dependance of 
the interstellar extinction on wavelength, the red region is more suitable for the study of the symbiotic sample, 
which has a spatial distribution flattened toward the Galactic plane.
 
However, the dominant reasons to move toward these red wavelengths lie in their high diagnostic
potential of the nature of the cool giant and the reduced disturbing presence of the nebular
continuum. Furthermore, this region (precisely the 8480-8740 \AA\ interval) has been selected for the
spectral survey ($3\times10^8$ objects, complete to $V\sim 17.5$ mag) that the ESA astrometric Cornerstone
mission GAIA should perform starting 2010.

The region is dominated by some of the strongest features visible in cool stars ( CaII triplet,
FeI, TiI, MgI and SiI lines, as well as lines from CN and TiO bands ) . The presence and activity 
of the hot companion is well traced by 
\clearpage
\begin{figure}
\plotfiddle{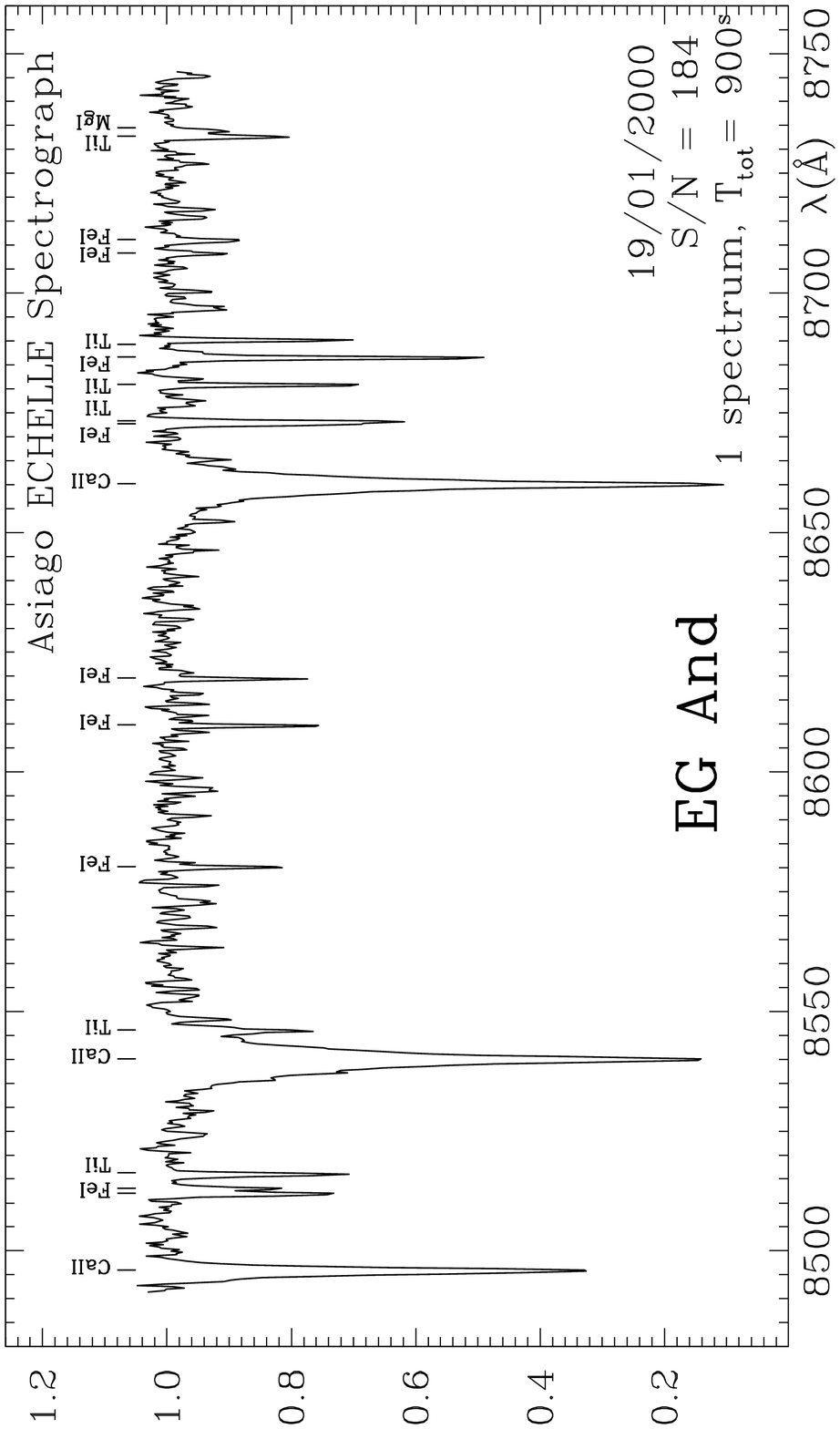}{19 truecm}{0}{80}{80}{-285}{-40}
\caption{Normalized spectrum of the symbiotic star EG~And over the GAIA wavelength region
($\lambda\lambda$ 8480-8740 \AA). No line appears in emission
over the well developped absorption continuum of the M giant (strongest lines belonging to
CaII, FeI, TiI and MgI, with many weaker lines contributed by CN and TiO molecules).}
\end{figure}
\begin{figure}
\plotfiddle{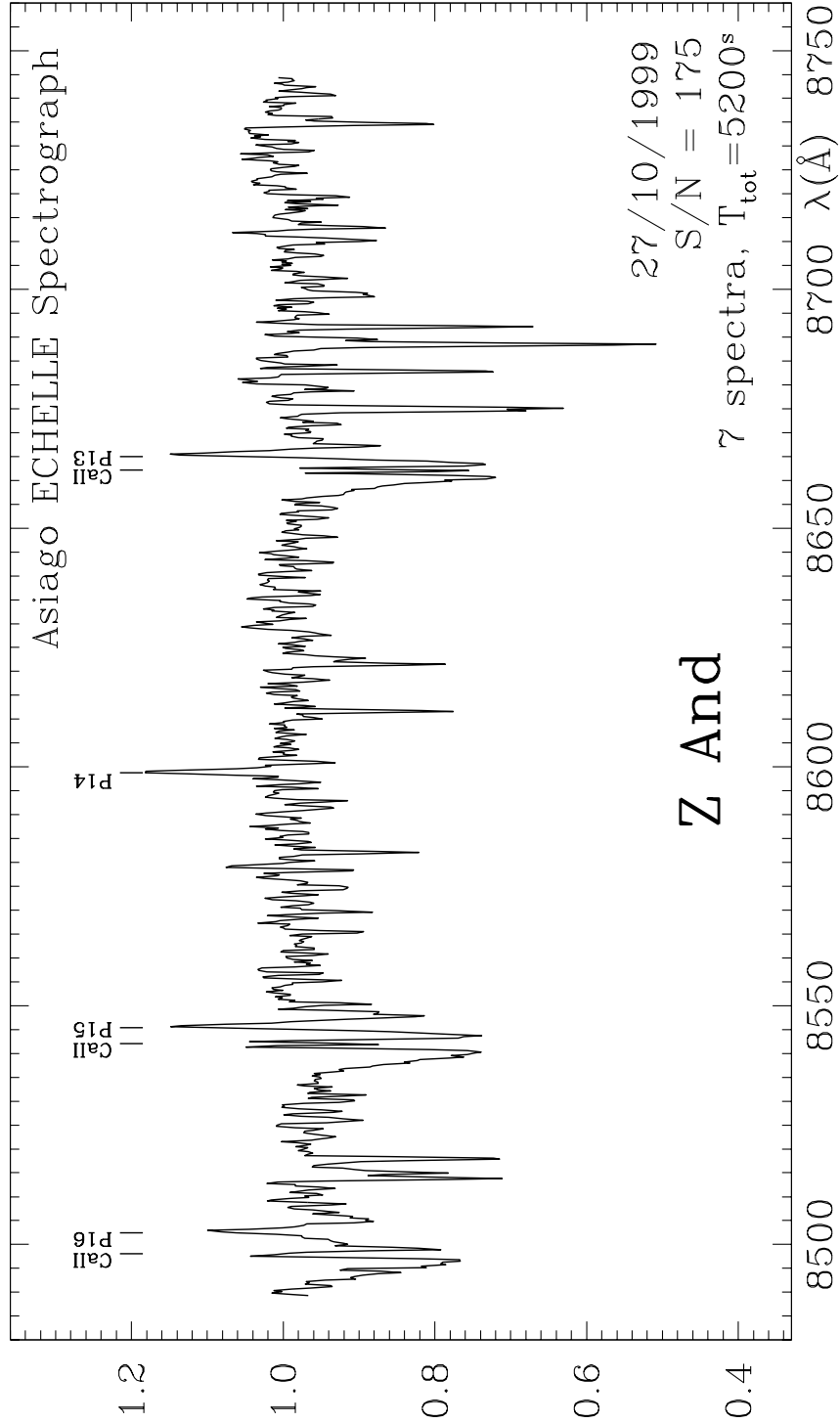}{19 truecm}{0}{80}{80}{-285}{-40}
\caption{Normalized spectrum of the symbiotic star Z~And over the GAIA wavelength region
($\lambda\lambda$ 8480-8740 \AA). Note the presence of double peaked emission core in the
CaII-triplet absorption lines and the P13, P14, P15 and P16 Paschen lines in emission. The
absorption spectrum of the M giant is undisturbed and rich in FeI, TiI and MgI absorptions
(apart from the forest of weak lines due to CN and TiO molecules).}
\end{figure}
\clearpage
\noindent
Paschen, CaII, HeI and NI emission lines which are strong and abundant
in this
wavelength interval. The latter is away from the veiling effect of the nebular blue continuum, 
that by filling-in aborption features tendes to mimic hotter spectral types, reduced metallicities 
and higher gravities for the cool component.
\section{The atlas}
We are completing a wide ranging, high resolution (R$\sim$20,000), high S/N ($\geq$100) spectral 
survey of symbiotic stars (selected from the Allen 1984 and Belczy\`nski et al. 2000 catalogues) 
over the 8480-8740 \AA~ wavelength range with the Asiago 1.82 telescope equipped with the Echelle+CCD 
spectrograph. Our collection of symbiotic spectra includes so far $\sim$40 objects, representative 
of the different types of cool components encountered in these binary systems (G, K, M, C and S spectral types, 
both non-variable as well as pulsating). Two examples of the collected spectra are presented in Figures~1 and 2.

The survey is supplemented by identical observations of a wide sample ($\sim$140) of MKK templates to serve as 
reference stars for classification purposes and training of syntetic model spectra aimed to investigate the cool 
component of symbiotic stars. The sample includes giants, bright giants and supergiants (ranging from G0 to M7, 
containing also C, S and Ba stars) and covers different metallicities ($-$2.59$\leq$[Fe/H]$\leq$+0.38). It is 
intended to integrate and enrich the GAIA atlas mapping the MKK system presented by Munari \& Tomasella (1999).
\section{Purposes}
The collected data will serve the purpose to ({\sl a}) classify the cool component with the greatest attention given 
to the determination of the still basically unknown luminosity class, ({\sl b}) derive their atmosferic parameters 
(T$_{{\sl eff}}$, log{\sl g}, [Fe/H], [{\sl $\alpha$}/Fe], V$_{{\sl rot}}$sin{\sl i}), ({\sl c}) asses the radial 
velocity dispersion of the symbiotic stars with respect to the standard galactic disk rotation and ({\sl d}) derive 
distances via spectrophotometric parallaxes, as well as ({\sl e}) understand better the kinematics and physical
properties of the nebular regions as mapped by the intensity and profile of the emission lines.

Knowing these properties would open unprecedented possibilities in addressing the evolutionary status of the symbiotic stars, 
their galactic population, their number in the Galaxy and the probability that they are precursors of type Ia supernovae.

\end{document}